\documentclass[%
 aip,
 amsmath,amssymb,
 reprint,%
]{revtex4-1}

\usepackage{graphicx}% Include figure files
\usepackage{dcolumn}% Align table columns on decimal point
\usepackage{bm}% bold math
\usepackage{amsmath}
\usepackage[utf8]{inputenc}
\usepackage[T1]{fontenc}
\usepackage{mathptmx}
\usepackage{etoolbox}
\usepackage[colorlinks=true,linkcolor=blue,citecolor=blue]{hyperref}%
\newcommand*{\citen}[1]{%
  \begingroup
    \romannumeral-`\x % remove space at the beginning of \setcitestyle
    \setcitestyle{numbers}%
    \cite{#1}%
  \endgroup   
}
\makeatletter
\def\@email#1#2{%
 \endgroup
 \patchcmd{\titleblock@produce}
  {\frontmatter@RRAPformat}
  {\frontmatter@RRAPformat{\produce@RRAP{*#1\href{mailto:#2}{#2}}}\frontmatter@RRAPformat}
  {}{}
}%
\makeatother

\begin{document}

\title{Gate-defined quantum point contacts in a germanium quantum well}

%\begin{comment}
\author{Han Gao}
 \affiliation{Beijing Key Laboratory of Quantum Devices, Key Laboratory for the Physics and Chemistry of Nanodevices, and School of Electronics, Peking University, Beijing 100871, China}
 
\author{Zhen-Zhen Kong}
 \affiliation{Key Laboratory of Microelectronics Devices $\&$ Integrated Technology, Institute of Microelectronics, Chinese Academy of Sciences, Beijing 100029, China}
 %\affiliation{Institute of Microelectronics, University of Chinese Academy of Sciences, Beijing 100049, China}

\author{Po Zhang}%
 \affiliation{Beijing Academy of Quantum Information Sciences, Beijing 100193, China}
 
\author{Yi Luo}%
 \affiliation{Beijing Key Laboratory of Quantum Devices, Key Laboratory for the Physics and Chemistry of Nanodevices, and School of Electronics, Peking University, Beijing 100871, China}
 \affiliation{Institute of Condensed Matter and Material Physics, School of Physics, Peking University, Beijing 100871, China}

\author{Haitian Su}%
 \affiliation{Beijing Key Laboratory of Quantum Devices, Key Laboratory for the Physics and Chemistry of Nanodevices, and School of Electronics, Peking University, Beijing 100871, China}
 \affiliation{Institute of Condensed Matter and Material Physics, School of Physics, Peking University, Beijing 100871, China}

\author{Xiao-Fei Liu}%
 \affiliation{Beijing Academy of Quantum Information Sciences, Beijing 100193, China}

\author{Gui-Lei Wang}
 \affiliation{Key Laboratory of Microelectronics Devices $\&$ Integrated Technology, Institute of Microelectronics, Chinese Academy of Sciences, Beijing 100029, China}
 \affiliation{Hefei National Laboratory, University of Science and Technology of China, Hefei, Anhui 230088, China}
 \affiliation{Beijing Superstring Academy of Memory Technology, Beijing 100176, China}
 \email{Guilei.Wang@bjsamt.org.cn}

\author{Ji-Yin Wang}%
 \affiliation{Beijing Academy of Quantum Information Sciences, Beijing 100193, China}
 \email{wang\_jy@baqis.ac.cn}
 
\author{H. Q. Xu}
 \affiliation{Beijing Key Laboratory of Quantum Devices, Key Laboratory for the Physics and Chemistry of Nanodevices, and School of Electronics, Peking University, Beijing 100871, China}
 \affiliation{Beijing Academy of Quantum Information Sciences, Beijing 100193, China}
 \email{hqxu@pku.edu.cn}
%\end{comment}


\begin{abstract}
 We report an experimental study of quantum point contacts defined in a high-quality strained germanium quantum well with layered electric gates. At zero magnetic field, we observe quantized conductance plateaus in units of 2$e^2/h$. Bias-spectroscopy measurements reveal that the energy spacing between successive one-dimensional subbands ranges from 1.5 to 5\,meV as a consequence of the small effective mass of the holes and the narrow gate constrictions. At finite magnetic fields perpendicular to the device plane, the edges of the conductance plateaus get splitted due to the Zeeman effect and Land\'{e} $g$ factors are estimated to be $\sim6.6$ for the holes in the germanium quantum well. We demonstrate that all quantum point contacts in the same device have comparable performances, indicating a reliable and reproducible device fabrication process. Thus, our work lays a foundation for investigating multiple forefronts of physics in germanium-based quantum devices that require quantum point contacts as a building block.      
       
\end{abstract}

\maketitle
Germanium (Ge) nanostructures are emerging as a pioneering research platform for pursuing multiple quantum computation schemes.\cite{Scappucci2021_NRM} This relies on the special properties in the holes of Ge, including strong spin-orbit interaction\cite{Bulaev2007_PRL,Moriya2014_PRL} and compatibility with superconducting components.\cite{Li2018_NL,Tosato2023_CM} Benefit from these properties, fast qubit operations in spin qubit processors\cite{Watzinger2018_NC,Hendrickx2020_NC,Hendrickx2020_Nature,Hendrickx2021_Nature,Froning2021_NN,Jirovec2021_NM,Jirovec2022_PRL,Wang2022_NC} and high-quality semiconductor-superconductor hybrids\cite{Ridderbos2020_NL,Tosato2023_CM} have been experimentally achieved in Ge nanostructures. Among the nanostructures, strained Ge quantum wells with two-dimensional hole gases are particularly attractive due to their flexibility in device preparation\cite{Hendrickx2021_Nature} and capability in large-scale integration.\cite{Borsoi2023_NN} On top of this, strained Ge quantum wells can be grown with hole mobility exceeding one million,\cite{Lodari2022_APL,Kong2023_AMI} signifying its exceptionally low disorder and long coherence length. The high quality of this material would lead to weak charge noise influence in quantum devices and therefore guarantees a high performance of the hole spin qubits.\cite{Hendrickx2020_Nature,Hendrickx2021_Nature} Aside from spin qubits, high-mobility Ge quantum wells are also considered as a competing platform for non-Abelian Majorana zero modes\cite{Adelsberger2023_PRB,Luethi2023_PRB,Laubscher2023_arxiv} and pioneering work has been done on hybridizing Ge quantum wells with superconductors.\cite{Hendrickx2018_NC,Hendrickx2019_PRB,Vigneau2019_NL,Aggarwal2021_PRR,Tosato2023_CM,Valentini2023_arxiv} 

Quantum point contacts (QPCs), as a basic nanostructure, have been widely used in quantum devices, including tunnelling barriers,\cite{Fornieri2019_Nature} quantum dots,\cite{vanderWiel2002_RMP,Hanson2007_RMP} charge sensors\cite{Elzerman2004_Nature} and fractional quantum Hall state interferometers.\cite{Bartolomei2020_Science,Nakamura2020_NP,Nakamura2023_PRX} In general, QPCs feature quantized conductance as a consequence of ballistic charge transport through one-dimensional (1D) channels.\cite{vanWees1988_PRL,Wharam1988_PRL} Previously, QPCs have been mainly studied in III-V materials, such as GaAs heterostructures,\cite{vanWees1988_PRL,Wharam1988_PRL} InAs quantum wells\cite{Debray2009_NN,Mittag2019_PRB,Lee2019_NL,Hsueh2022_PRB} and InSb quantum wells.\cite{Qu2016_NL,Lei2021_PRR} In contrast, QPCs in strained Ge quantum wells are rarely studied.\cite{Gul2017_APL,Mizokuchi2018_NL,Gul2018_JPCM} In the mean time, QPCs are a key component in aforementioned Ge-based quantum devices including spin qubit processors and semiconductor-superconductor hybrid devices. Therefore, QPCs in strained Ge quantum wells deserve to be investigated thoroughly, especially from their reproducibility perspective.

In this work, we have fabricated QPCs in a strained Ge quantum well and studied their physical properties by electrical transport measurements. The QPCs in the Ge quantum well are defined by layered electric gates. At zero magnetic field, the QPCs exhibit conductance quantization in units of $g_{Q}=2e^2/h$, indicating a ballistic charge transport through the gate-defined 1D channels. Bias-spectroscopy measurements are then conducted and quantization energies between the 1D subbands are obtained with values from 1.5 to 5$\,\mathrm{meV}$. The considerable quantization energies result from the small effective mass of the holes in Ge and the narrow gate constrictions in the device. With magnetic fields, the edges of the conductance plateaus get splitted due to the Zeeman effect and the Zeeman energies are quantified in bias-spectroscopy measurements. Then, Land\'{e} $g$ factors of the 1D subbands are estimated to be $\sim6.6$. Notably, all three QPCs fabricated in the same device show comparable behaviors as a confirmation of uniform device preparations. Thus, our work demonstrates a reliable way of constructing quantum devices in Ge quantum wells.

\begin{figure*}[!t]
\centering
\includegraphics[width=0.93\linewidth]{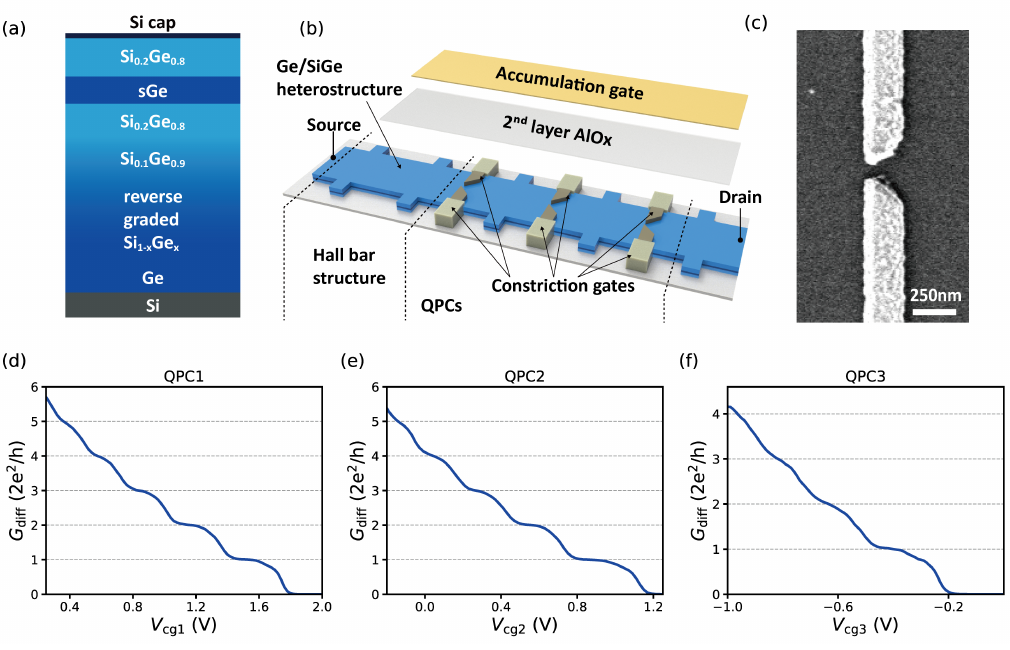}
\caption{(a) Layer schematic of the Ge/SiGe heterostructure. (b) Schematic sketch of a device with three QPCs and a Hall bar structure. Each QPC is defined by a pair of constriction gates on the Hall-bar mesa structure. There is a 25-nm AlOx dielectric layer between the constriction gates and the Ge/SiGe heterostructure. Another layer of 25-nm AlOx is grown on top of the constriction gates. A global accumulation gate on top of the device extends to covering contact regions including source and drain. (c) SEM image of a representative QPC structure before depositing accumulation gate. The separation between the pair of constriction gates is $\sim 100\,\mathrm{nm}$. (d)-(f) Zero-bias differential conductance $G_\text{diff}$ as a function of corresponding constriction gates $V_\text{cgi}$ ($i$=1, 2, 3) for the three QPCs at $B=0\,\mathrm{T}$ and $V_\text{ag}=-10\,\mathrm{V}$. When one QPC is measured, the constriction gates of the other two QPCs are set to $-2.5\,\mathrm{V}$.
}\label{Figure1}
\end{figure*}

Figure \ref{Figure1}(a) shows the layer structure of the Ge quantum well used in this work which is grown by reduced pressure chemical vapor deposition (RPCVD).\cite{Kong2023_AMI} The Ge/SiGe heterostructure is cultivated on a Si(001) substrate where $1.7\,\mathrm{\mu m}$ Ge is initially grown. On top of this, there is a 1.31 $\mu$m SiGe layer with gradually varied Si and Ge concentrations. A 10-nm thick Ge quantum well is subsequently grown, followed by a growth of $34\,\mathrm{nm}$ Si$_{0.2}$Ge$_{0.8}$ as a top barrier layer. The heterostructure growth is accomplished with a $1.28\,\mathrm{nm}$ thick Si capping layer. More details of the heterostructure growth are described in Ref.\citen{Kong2023_AMI}. With the Ge quantum well, we have prepared a device as sketched in Figure \ref{Figure1}(b). The device fabrication starts with defining a Hall bar-shaped mesa using UV photolithography and reactive ion etching (RIE). The etched depth is $\sim120\,\mathrm{nm}$ and surpasses the depth of the Ge quantum well, which ensures the formation of conduction channels within the mesa exclusively. Then, contact leads are obtained by depositing $60\,\mathrm{nm}$ thick Pt, after buffered oxide etch (BOE) is utilized to remove the native oxide layer on the heterostructure surface. Afterwards, an annealing process at $350\,^\circ$C is taken in a high vacuum chamber, facilitating a good contact with the Ge quantum well. A $25\,\mathrm{nm}$ thick AlOx layer is grown via atomic layer deposition (ALD) at $150\,^\circ$C. The AlOx layer serves as a dielectric between the heterostructure and the successive electric gate layer. Then, three pairs of constriction gates are fabricated by pattern definition via electron-beam lithography, metal deposition of $5/25\,\mathrm{nm}$ Ti/Pd via electron-beam evaporation, and lift-off. As shown in the scanning electron microscope (SEM) image in Figure \ref{Figure1}(c), each pair of constriction gates is typically separated by $\sim100\,\mathrm{nm}$. A subsequent thick connection layer of Ti/Au ($10/190\,\mathrm{nm}$) is deposited to connect the constriction gates to bonding pads. A second layer of $25\,\mathrm{nm}$ AlOx is grown by ALD on top of the sample. Ultimately, a global accumulation gate is fabricated by pattern definition via UV photolithography, metal deposition of $10/300\,\mathrm{nm}$ Ti/Au, and lift-off. As seen in Figure \ref{Figure1}(b), the fabricated device is composed of a Hall bar structure and three QPCs, and the two parts share the same accumulation gate. This architecture ensures that the electrical properties of the Ge quantum well at every specific accumulation gate voltage can be obtained when demanded.       

\begin{figure}[!t]
\centering
\includegraphics[width=1\linewidth]{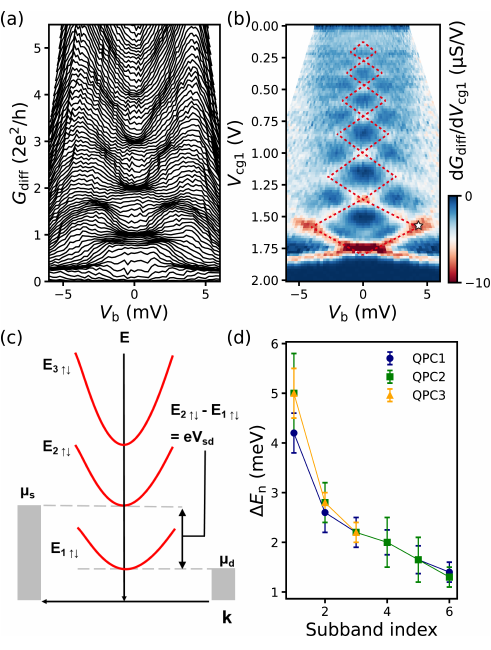}
\caption{(a) Waterfall plot of $G_\text{diff}$ as a function of $V_\text{b}$, at varied $V_\text{cg1}$ from $0\,\mathrm{V}$ to $2\,\mathrm{V}$ with a step of $0.02\,\mathrm{V}$ ($B=0\,\mathrm{T}$ and $V_\text{ag}=-10\,\mathrm{V}$). Dense regions correspond to conductance plateaus. (b) Transconductance d$G_\text{diff}$/d$V_\text{cg1}$ of the QPC1 as a function of $V_\text{b}$ and $V_\text{cg1}$. Red dashed lines, as a guide to the eye, enclose diamond shapes. (c) Energy diagram of the 1D subbands at the star point in panel (b). Each parabola corresponds to a spin-degenerated subband with index $n$ ($n=1,2,3...$). The energy spacing between the $n$th and $(n+1)$th subband is labelled as $\Delta E_\text{n}$. In the diagram, $\mu_\text{s}-\mu_\text{d}$ (or $eV_\text{b}$) just matches $\Delta E_\text{1}$. (d) Extracted energy spacing between subbands $\Delta E_\text{n}$ as a function of subband index for the three QPCs. $\Delta E_\text{n}$ of the three QPCs are obtained from 2D plots as shown in panel (b). The error bars correspond to the full width at half maximum of transconductance peaks.}\label{Figure2}
\end{figure}

The fabricated device is loaded into a He$^{4}$ cryostat and all electrical transport measurements are performed at $1.8\,\mathrm{K}$. The electrical measurements of the Hall bar section are accomplished with a standard lock-in setup (see Figure S1 in the Supplementary Materials). The measurements of the QPCs are done with a two-terminal setup, where the source and drain leads are connected to the measurement circuit (see Figure S2 in the Supplementary Materials). In the measurements, both ac and dc techniques are employed and the two signals are combined with a summing module before feeding into the device. A low-frequency ac excitation $V_\text{ac}$ is applied and corresponding ac current $I_\text{ac}$ are detected at varied dc bias voltages. In order to obtain the precise differential conductance of the QPCs, serial resistance $R_\text{s}$ arising from non-QPC sections needs to be accounted for. The differential conductance is calculated as $G_\text{diff}=1/(V_\text{ac}/I_\text{ac}-R_\text{s})$ and corresponding dc bias voltage is corrected accordingly by subtracting the voltage drop across $R_\text{s}$. More details about the corrections are described in the second section of the Supplementary Materials. All transport data shown in the main article have been subtracted by corresponding serial resistances. 

Before studying the QPCs, we first characterize the Hall bar section with magnetoresistance measurements. Carrier density $p$ and hole mobility $\mu_\text{h}$ as a function of the accumulate gate $V_\text{ag}$ are studied (see Figure S3 in the Supplementary Materials). At $V_\text{ag}=-10\,\mathrm{V}$, the carrier density $p$ is $2.3\times10^{11}\mathrm{cm^{-2}}$ and the hole mobility $\mu_\text{h}$ is $3.5\times10^{5}\mathrm{cm^2}/\mathrm{(V\ s)}$. With the above parameters, the mean free path of the holes is estimated to be $l_\text{m}\sim$2.8 $\mathrm{\mu}\mathrm{m}$, significantly larger than the size of the constriction gates. Then, ballistic transport should be allowed for the QPCs at $V_\text{ag}=-10\,\mathrm{V}$ and thus $V_\text{ag}$ is fixed at this value in all QPC measurements. Figures \ref{Figure1}(d)-\ref{Figure1}(f) show zero-bias $G_\text{diff}$ as a function of corresponding constriction gates at $B=0\,\mathrm{T}$. When one QPC is being measured, the constriction gates of the other two QPCs are set to $-2.5\,\mathrm{V}$ in order to open channels under the two QPCs. Conductance plateaus up to at least 3 times $g_{Q}$ are discernible for all three QPCs, where $g_{Q}=2e^2/h$. Particularly, QPC1 and QPC2 have visible conductance plateaus up to 5 times $g_{Q}$. The conductance plateaus at quantized values result from charge transport through 1D subbands defined by the constriction gates. Aside from the conductance plateaus at integer multiples of $g_{Q}$, we observe conductance shoulders at $G_\text{diff}\sim0.7g_{Q}$ in Figures \ref{Figure1}(d)-\ref{Figure1}(f). Such anomalous conductance shoulders have been thoroughly investigated in a previous work and can be ascribed to electron-electron interaction.\cite{Cronenwett2002_PRL} 

\begin{figure*}[!t]
\centering
\includegraphics[width=0.93\linewidth]{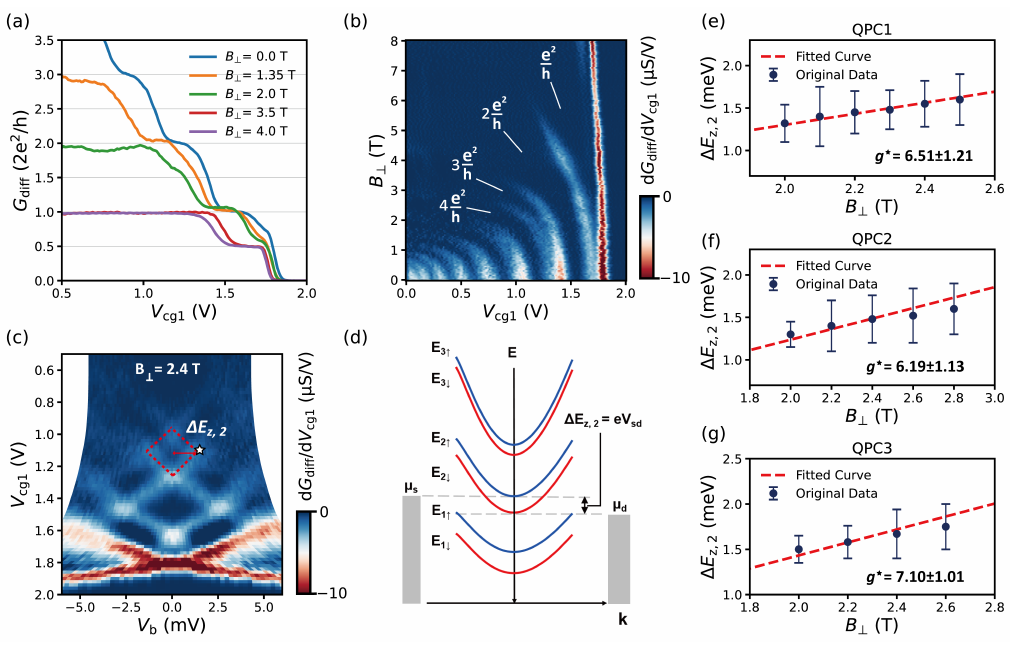}
\caption{(a) $G_\text{diff}$ of QPC1 as a function of $V_\text{cg1}$ at several different perpendicular magnetic fields $B_{\bot}$ varied from 0 to $4\,\mathrm{T}$. Half-integer plateaus become visible at high $B_{\bot}$ due to the Zeeman effect. (b) Transconductance d$G_\text{diff}$/d$V_\text{cg1}$ of QPC1 as a function of $B_{\bot}$ and $V_\text{cg1}$ at $V_{b}=0$. The red, bright fan features seen in the figure represent the evolutions of the conductance plateau edges with increasing $B_{\bot}$. The conductance plateau regions are labelled by their corresponding quantized conductance values in the figure. (c) Transconductance d$G_\text{diff}$/d$V_\text{cg1}$ of QPC1 as a function of $V_\text{b}$ and $V_\text{cg1}$ at $B_{\bot}=2.4\,\mathrm{T}$. (d) Energy diagram of the spin-splitted subbands at finite $V_\text{b}$ as indicated by the star point in panel (c). In the diagram, $\mu_\text{s}-\mu_\text{d}$ (or $eV_\text{b}$) just matches $\Delta E_\text{z,2}$. (e)-(g) Zeeman splittings of the second subband $\Delta E_\text{z,2}$ as a function of $B_{\bot}$ for the three QPCs. The blue points are extracted from 2D maps as in panel (c). The error bars correspond to the full width at half maximum of transconductance peaks. Red dashed lines are linear fits to the blues points and obtained $g^{\star}$ of the subbands are shown in the panels.
}\label{Figure3}
\end{figure*}

In the next, bias-spectroscopy measurements are employed to gain further insight into the energy scale of the gate-defined 1D subbands in the QPCs. In Figure \ref{Figure2}, we perform bias-spectroscopy measurements on the QPCs in the absence of magnetic field and quantization energies between the 1D subbands are obtained. Figure \ref{Figure2}(a) shows a waterfall plot of $G_\text{diff}$ as a function of $V_\text{b}$ with varied $V_\text{cg1}$ for QPC1. In the plot, $V_\text{cg1}$ is varied from $0\,\mathrm{V}$ to $2\,\mathrm{V}$ with a step of $0.02\,\mathrm{V}$ and there is no offset between the traces. Dense regions correspond to conductance plateaus. At low $V_\text{b}$, dense regions appear at integer multiples of $g_{Q}$ as well as at $\sim0.7g_{Q}$. Half integer plateaus emerge at large $V_\text{b}$ when more than one subbands drops into the bias voltage window.\cite{Kouwenhoven1989_PRB,Weperen2013_NL} Then, we derive the conductance $G_\text{diff}$ with respect to the constriction gate voltage $V_\text{cg1}$ and obtain transconductance d$G_\text{diff}$/d$V_\text{cg1}$ as a function of $V_\text{b}$ and $V_\text{cg1}$ [see Figure \ref{Figure2}(b)]. In the figure, red dashed lines denote six consecutive diamonds implying that the first six 1D subbands are well separated in energy. Another set of faint diamond appears inside the bottom diamond which is a result of the $0.7g_{Q}$ anomaly.\cite{Kouwenhoven1989_PRB} Then, we perform analogous bias-spectroscopy measurements on QPC2 and QPC3 as shown in Figure S4 in the Supplementary Materials. With such spectroscopy measurements, quantization energies of 1D subbands can be extracted. Figure \ref{Figure2}(c) shows the energy diagram of the subbands at a finite $V_\text{b}$ [star point in Figure \ref{Figure2}(b)]. Each red parabola corresponds to a spin-degenerated 1D subband with subband index $n$ ($n=1, 2, 3...$). The chemical potential difference $\mu_\text{s}-\mu_\text{d}$ (also $eV_\text{b}$) just matches the energy spacing between the first and second subband, denoted as $\Delta E_\text{1}$, and the energy spacing is then extracted. Similarly, energy spacing $\Delta E_\text{n}$ depending on index $n$ is obtained for the three QPCs and corresponding results are displayed in Figure \ref{Figure2}(d). We see that $\Delta E_\text{n}$ decreases with $n$, which can be explained by a modulated QPC confinement potential profile with constriction gates. Populating more subbands needs more negative $V_\text{cg}$ and thus reduces the curvature of the confinement potential, leading to smaller quantization energies of the subbands with larger indices. By adopting a harmonic potential for constriction gate confinement, the extension length of the confinement for a subband can be estimated as \cite{Hsueh2022_PRB} \begin{eqnarray}
\frac{1}{2}m^{*}\omega^{2}_\text{0}L^{2}_\text{n}=\hbar\omega_\text{0}(n-\frac{1}{2}),\label{eq:one}
\end{eqnarray} where $m^{*}=0.1m_\text{e}$\cite{Kong2023_AMI} is the effective mass of the holes, $\omega_\text{0}$ is the angular frequency, and $L_\text{n}$ is the effective length of the lateral confinement for the $n$th subband. With Eq.~(\ref{eq:one}), the $L_\text{n}$ values for the first six subbands of QPC1 is estimated to be 13, 30, 42, 52, 64 and 77$\,\mathrm{nm}$, respectively. The values of $L_\text{n}$ for the other two QPCs are provided in Figure S4 in the Supplementary Materials. Thus, the effective length $L_\text{n}$ of the first several subbands of the QPCs ranges from 12 to 80$\,\mathrm{nm}$, smaller than the physical separations between the constriction gates in pairs ($\sim100\,\mathrm{nm}$). Notably, the three QPCs have similar quantization energies as a function of the subband index, indicating a comparable gate confinement and a good reproducibility in device fabrication. 

In the following, we focus on spin dependent behaviors of the QPCs subjected to perpendicular magnetic fields $B_{\bot}$. Figure \ref{Figure3}(a) displays zero-bias $G_\text{diff}$ of QPC1 as a function of $V_\text{cg1}$ at several selected $B_{\bot}$. At $B=0$ (blue trace), conductance plateaus appear mainly at integer multiples of $g_{Q}$. Due to the Zeeman effect, conductance shoulders at half integer of $g_{Q}$ emerge from $B_{\bot}=1.35\,\mathrm{T}$ and the half integer plateaus become well visible at $B_{\bot}=4\,\mathrm{T}$. Figure \ref{Figure3}(b) presents transconductance d$G_\text{diff}$/d$V_\text{cg1}$ as a function of $V_\text{cg1}$ and $B_{\bot}$. A large absolute value of d$G_\text{diff}$/d$V_\text{cg1}$ represents that a spin-degenerate or spin-indegenerate subband gets populated with decreasing the constriction gate voltage $V_\text{cg1}$. With increasing $B_{\bot}$, we see that the subbands get splitted in energy and spin-indegenerate subbands are obtained. Except for the Zeeman effect, $B_{\bot}$ has another influence that all subbands bend to low gate voltages due to the orbital effect\cite{Lei2021_PRR,Mizokuchi2018_NL,TranInNano_book2009}. This influence is more prominent for high-index subbands and these subbands are depopulated by the magnetic field even before a visible Zeeman splitting commences. Therefore, Zeeman effect on the QPCs is only observable for low-index subbands. In order to obtain Zeeman energies of the subbands, we conduct bias-spectroscopy measurements on QPCs at finite $B_{\bot}$. Figure \ref{Figure3}(c) shows d$G_\text{diff}$/d$V_\text{cg1}$ as a function of $V_{b}$ and $V_\text{cg1}$ at $B_{\bot}=2.4\,\mathrm{T}$. Compared with Figure \ref{Figure2}(b) ($B=0$), spin-splitted subbands become visible in Figure \ref{Figure3}(c). Red dashed lines denote the boundary of the diamond formed by the second subband with a finite Zeeman energy. Then, the Zeeman energy of the second subband $\Delta E_\text{z,2}$ can be extracted at $B_{\bot}=2.4\,\mathrm{T}$. Figure \ref{Figure3}(d) depicts the energy diagram of the spin-splitted subbands at finite $V_\text{b}$ [star point in Figure \ref{Figure3}(c)]. Blue parabolas correspond to subbands with spin-up species and red parabolas are for spin-down subbands. Here, the chemical potential difference $\mu_\text{s}-\mu_\text{d}$ (also $eV_\text{b}$) just equals the Zeeman splitting of the second subband $\Delta E_\text{z,2}$. A series of measurements as in Figure \ref{Figure3}(c) at different $B_{\bot}$ are performed for QPC1 (see Figure S5 in the Supplementary Materials) and Zeeman energies extracted at different $B_{\bot}$ are shown in Figure \ref{Figure3}(e). Similar measurements have been done on QPC2 and QPC3, and corresponding results are shown in Figure S6 and S7 in the Supplementary Materials. Extracted Zeeman splittings of QPC2 and QPC3 as a function of $B_{\bot}$ are displayed in Figure \ref{Figure3}(f) and \ref{Figure3}(g), respectively. In Figures \ref{Figure3}(e)-\ref{Figure3}(g), we make linear fits to the data (red dashed lines) and Land\'{e} $g$ factors of the second subband for the three QPCs are obtained. The extracted Land\'{e} $g$ factors are $\sim6.6$, consistent with previous reports on Ge hole gases.\cite{Sammak2019_AFM,Mizokuchi2018_NL,Kong2023_AMI} Note that we only deal with Zeeman energy of the second subband because Zeeman splitting of the first subband is influenced by the $0.7g_{Q}$ anomaly (see Figure S5-S7) and higher subbands do not have visible splitting before being fully depopulated by $B_{\bot}$.

In conclusion, we have experimentally studied QPCs defined in a Ge quantum well with electric gates. At zero magnetic field, we observe conductance plateaus at quantized values for all three QPCs, indicating a ballistic charge transport through 1D subbands. Bias-spectroscopy measurements are employed on the QPCs to obtain quantization energies between the subbands. The value varies with subband index from 1.5 to 5$\,\mathrm{meV}$ as a consequence of modulated gate confinements. After that, the QPCs are studied in perpendicular magnetic fields and we observe splitted subbands in energy due to the Zeeman effect. With the extracted Zeeman splitting at different magnetic fields, Land\'{e} $g$ factors are obtained to be $\sim6.6$. Importantly, all three QPCs in the same device have uniform performances with regard to their lateral gate confinements and Land\'{e} $g$ factors. The uniformity is a key issue for constructing complex quantum devices with basic nanostructures. We therefore believe that our work is essentially important for pursuing Ge-based quantum technology in devices using QPC structures as a building block. 

%The fabricated device is composed of three QPCs and a Hall bar structure. Three QPCs in one device helps to gain insight into device quality with regard to it reproducibility and the co-fabricated Hall bar is used to learn the basic electrical properties of the Ge quantum well. The considerable quantization energies arise from the small effective mass of the holes and the strong confinement from the constriction gates.%
\begin{acknowledgments}
This work is supported by the National Natural Science Foundation of China (Grants No. 12374480, 92165208, 11874071 and 92365103), and Innovation Program for Quantum Science and Technology (Project ID. 2021ZD0302301).
\end{acknowledgments}

\section*{conflict of interest}
The authors declare no conflict of interests. 

\section*{data availability}
The data that support the findings of this study are available from
the corresponding authors upon reasonable request.

\bibliography{reference}% Produces the bibliography via BibTeX.
\end{document}

\title{Supplementary Materials: Gate-defined quantum point contacts in a germanium quantum well}
%\begin{comment}
\author{Han Gao}
 \affiliation{Beijing Key Laboratory of Quantum Devices, Key Laboratory for the Physics and Chemistry of Nanodevices, and School of Electronics, Peking University, Beijing 100871, China}
 
\author{Zhen-Zhen Kong}
 \affiliation{Key Laboratory of Microelectronics Devices $\&$ Integrated Technology, Institute of Microelectronics, Chinese Academy of Sciences, Beijing 100029, China}
 %\affiliation{Institute of Microelectronics, University of Chinese Academy of Sciences, Beijing 100049, China}

\author{Po Zhang}%
 \affiliation{Beijing Academy of Quantum Information Sciences, Beijing 100193, China}
 
\author{Yi Luo}%
 \affiliation{Beijing Key Laboratory of Quantum Devices, Key Laboratory for the Physics and Chemistry of Nanodevices, and School of Electronics, Peking University, Beijing 100871, China}
\affiliation{Institute of Condensed Matter and Material Physics, School of Physics, Peking University, Beijing 100871, China}

\author{Haitian Su}%
 \affiliation{Beijing Key Laboratory of Quantum Devices, Key Laboratory for the Physics and Chemistry of Nanodevices, and School of Electronics, Peking University, Beijing 100871, China}
\affiliation{Institute of Condensed Matter and Material Physics, School of Physics, Peking University, Beijing 100871, China}

\author{Xiao-Fei Liu}%
 \affiliation{Beijing Academy of Quantum Information Sciences, Beijing 100193, China}

\author{Gui-Lei Wang}
 \affiliation{Key Laboratory of Microelectronics Devices $\&$ Integrated Technology, Institute of Microelectronics, Chinese Academy of Sciences, Beijing 100029, China}
 \affiliation{Hefei National Laboratory, University of Science and Technology of China, Hefei, Anhui 230088, China}
 \affiliation{Beijing Superstring Academy of Memory Technology, Beijing 100176, China}
 \email{Guilei.Wang@bjsamt.org.cn}

\author{Ji-Yin Wang}%
 \affiliation{Beijing Academy of Quantum Information Sciences, Beijing 100193, China}
 \email{wang\_jy@baqis.ac.cn}
 
\author{H. Q. Xu}
 \affiliation{Beijing Key Laboratory of Quantum Devices, Key Laboratory for the Physics and Chemistry of Nanodevices, and School of Electronics, Peking University, Beijing 100871, China}
 \affiliation{Beijing Academy of Quantum Information Sciences, Beijing 100193, China}
 \email{hqxu@pku.edu.cn}
%\end{comment}


\maketitle
\section{Measurement setup and basic characterization}
In this section, we describe the measurement setups used in this work and the basic characteristics of the device. 

\begin{figure*}[!t]
\centering
\includegraphics[width=0.65\linewidth]{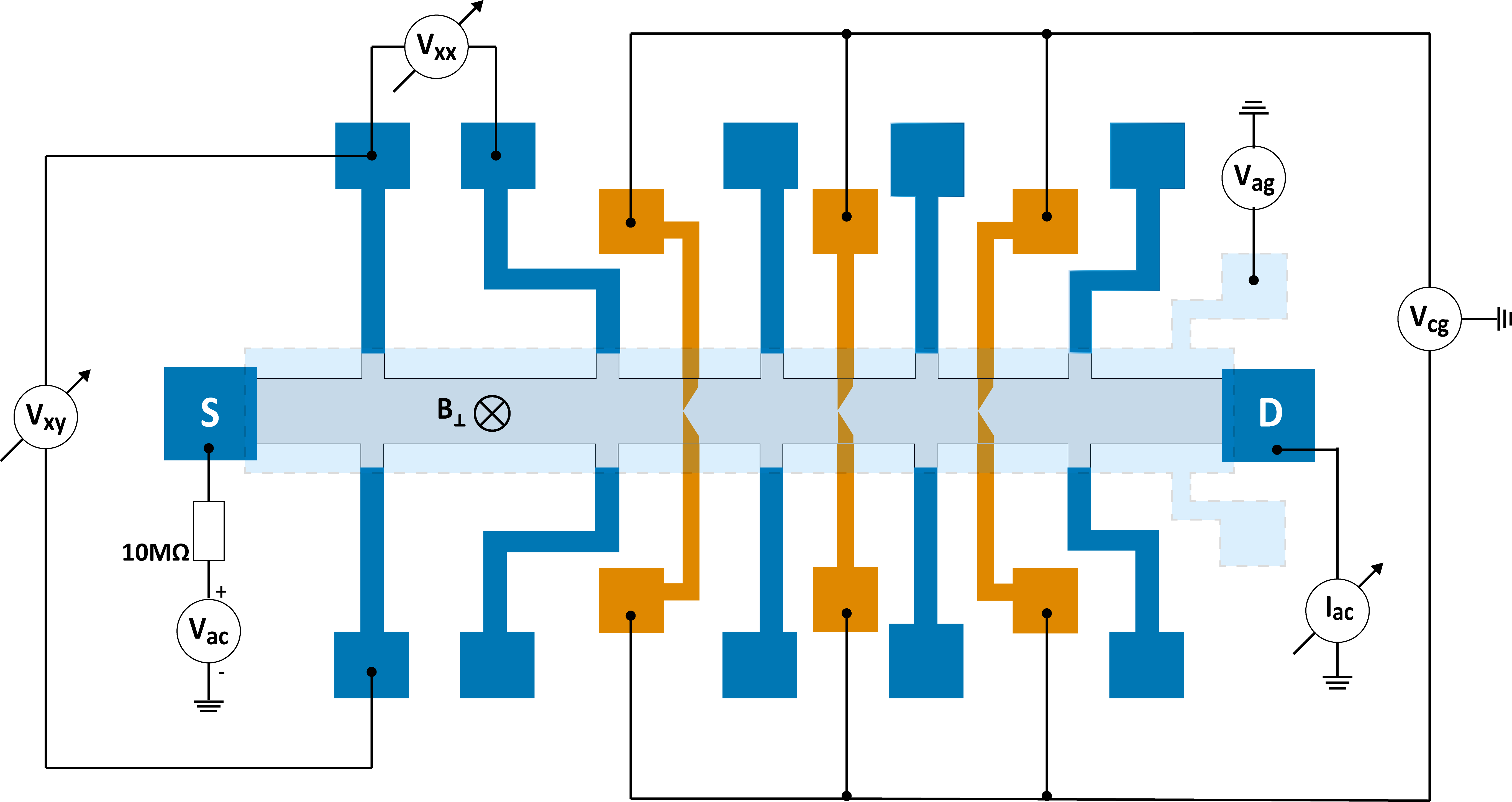}
\caption{Measurement setup for the Hall bar structure. An ac voltage $V_\text{ac}=500\,\mathrm{mV}$ is applied with a lock-in instrument to the source lead (S) and ac current $I_\text{ac}$ is measured from the drain lead (D). A 10 M$\Omega$ resistor is used to maintain an ac current of $\sim50\,\mathrm{nA}$. Longitudinal voltage $V_\text{xx}$ and transversal Hall voltage $V_\text{xy}$ are measured with lock-in instruments. All lock-in instruments are operated at a frequency of $17.77\,\mathrm{Hz}$. Contacts are colored in blue and five of them are used in the measurements while rest contacts are floated. The voltages of three pairs of constriction gates (orange color) are fixed at $V_\text{cg}=-2.5\,\mathrm{V}$ to open up the channel below the constriction gates. As a global gate, the accumulation gate (light blue) with voltage $V_\text{ag}$ is used to tune the carrier density in the whole device. A perpendicular magnetic field $B_{\bot}$ is employed during the Hall bar measurements.}\label{FigureS1}
\end{figure*}

Figure \ref{FigureS1} illustrates the measurement setup used for the Hall bar structure. Contact leads are colored in blue and five of them are used in the measurements. An ac voltage $V_\text{ac}=500\,\mathrm{mV}$ is applied onto a 10 M$\Omega$ resistor via a lock-in instrument before connecting to the source lead, which would lead to a $\sim50\,\mathrm{nA}$ ac current. In the mean time, ac current $I_\text{ac}$ is measured from the drain lead in order to detect the actual current running in the circuit. Longitudinal voltage $V_\text{xx}$ and transversal Hall voltage $V_\text{xy}$ are measured with lock-in instruments. All lock-in instruments are operated at a frequency of $17.77\,\mathrm{Hz}$. Three pairs of constrictions gates (orange) are applied with a voltage $V_\text{cg}=-2.5\,\mathrm{V}$ to open up the regions below the constriction gates. As a global gate, the accumulation gate (light blue) with voltage $V_\text{ag}$ is used to tune the carrier density in the whole device. A perpendicular magnetic field $B_{\bot}$ is applied during the Hall measurements. Then, we obtain carrier density $p$ and hole mobility $\mu_\text{h}$ of the Ge quantum well as a function of the accumulation gate $V_\text{ag}$ (see Figure \ref{FigureS3}). At $V_\text{ag}=-10\,\mathrm{V}$, the carrier density $p$ is $2.3\times10^{11}\mathrm{cm^{-2}}$ and the hole mobility $\mu_{h}$ is $3.5\times10^{5}\mathrm{cm^2}/\mathrm{(V\ s)}$. In all the QPC measurements reported in the present work, $V_\text{ag}$ is fixed at $-10\,\mathrm{V}$. 

\begin{figure*}[!t]
\centering
\includegraphics[width=0.65\linewidth]{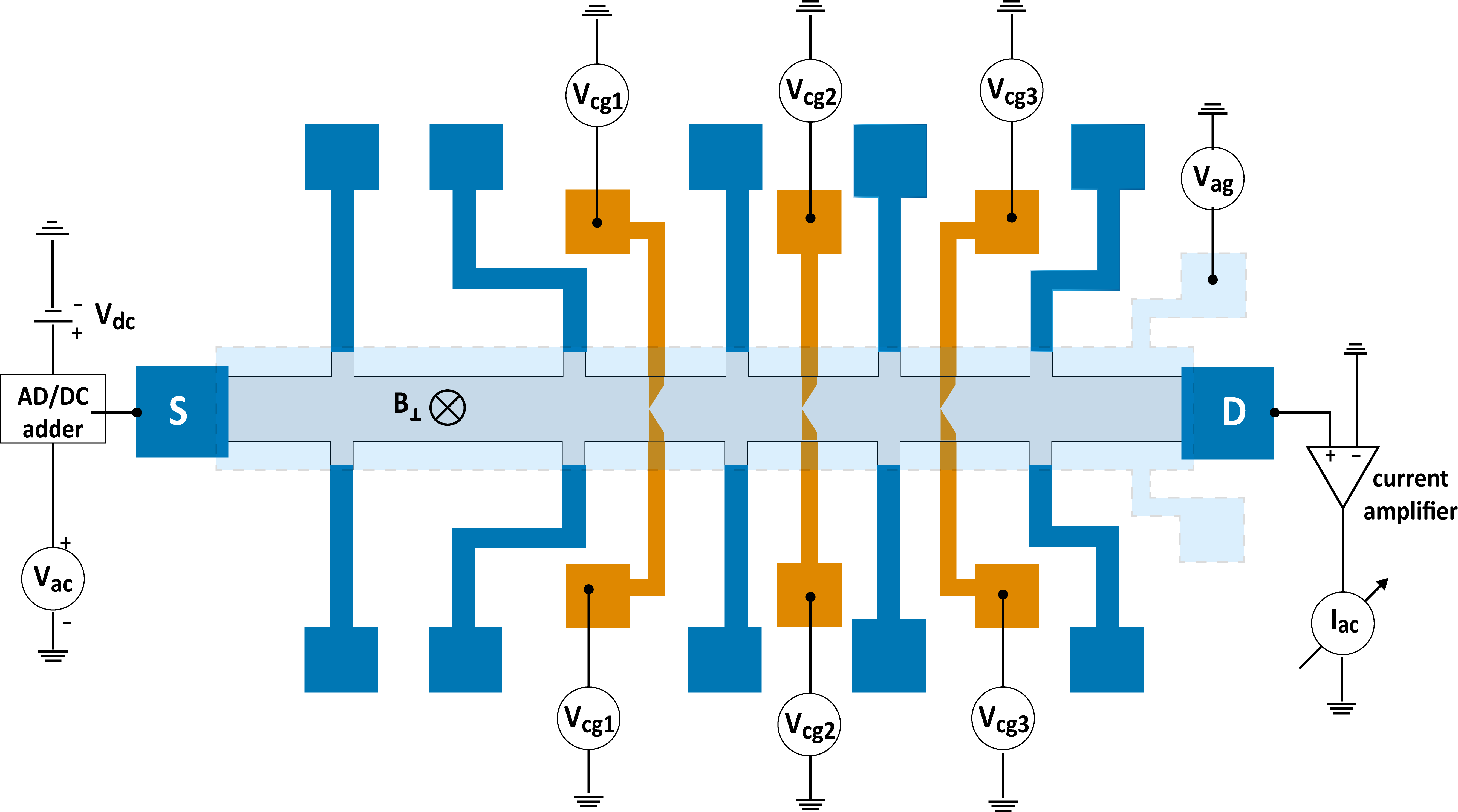}
\caption{Measurement setup for QPC measurements. An ac voltage $V_\text{ac}$ and a dc voltage $V_\text{dc}$ are summed up with a summing module before feeding into the source lead, while the ac current $I_\text{ac}$ is measured with the help of a current pre-amplifier. Three QPCs are defined with three pairs of constriction gates with voltage $V_\text{cg1}$, $V_\text{cg2}$ and $V_\text{cg3}$. The accumulation gate with voltage $V_\text{ag}$ is used to tune the carrier density in the device globally.}\label{FigureS2}
\end{figure*}

\begin{figure*}[!t]
\centering
\includegraphics[width=0.5\linewidth]{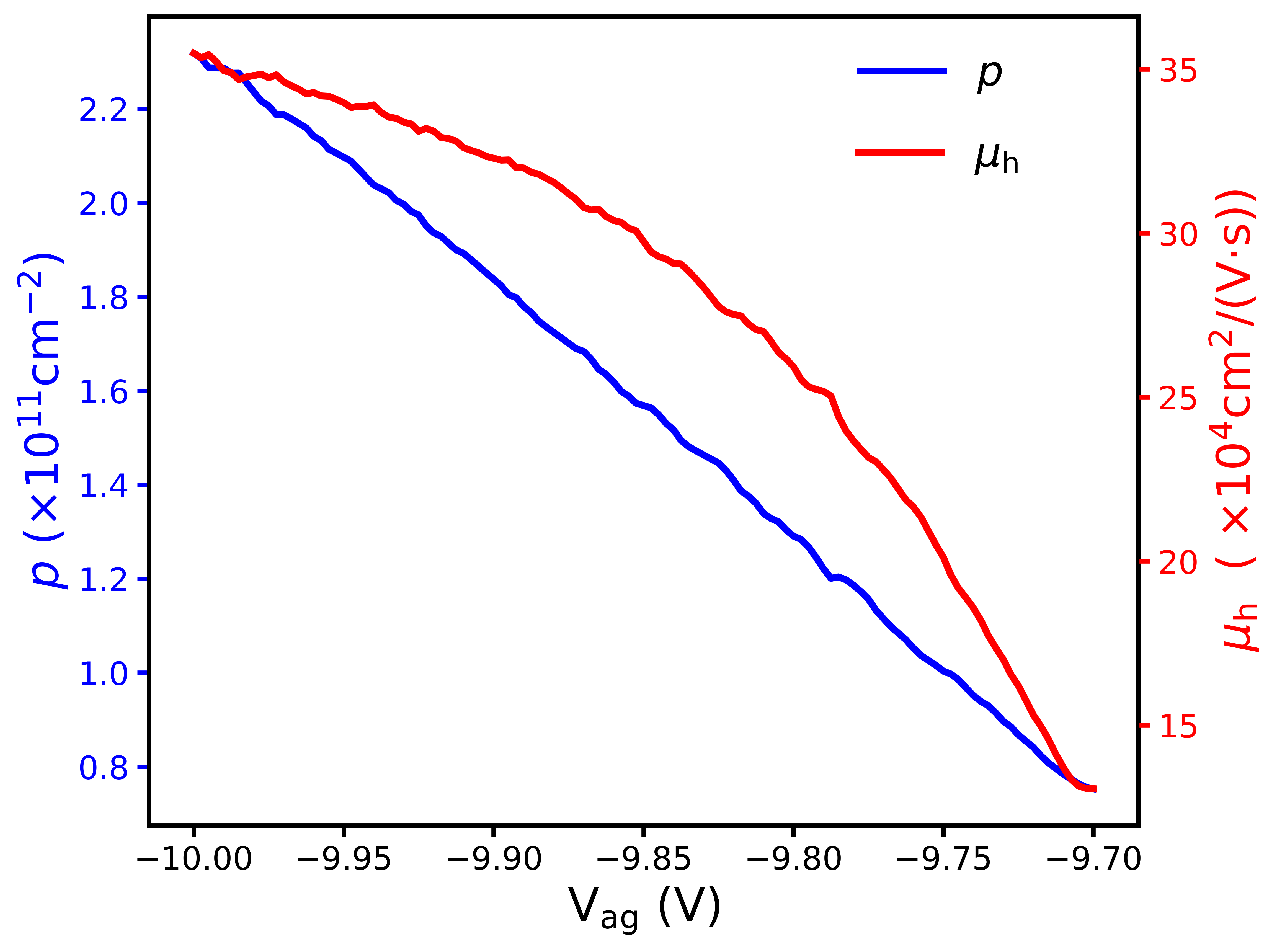}
\caption{Carrier density $p$ and hole mobility $\mu_\text{h}$ in the Ge quantum well as a function of the accumulation gate $V_\text{ag}$. Corresponding data are measured from the Hall bar structure with a measurement setup shown in Figure S1. The accumulation gate is fixed at $V_\text{ag}=-10\,\mathrm{V}$ in all the QPC measurements.}\label{FigureS3}
\end{figure*}

Figure \ref{FigureS2} displays the measurement setup for QPC measurements. An ac voltage $V_\text{ac}$ and a dc voltage $V_\text{dc}$ are summed up with a summing module before feeding into the source lead. An ac current $I_\text{ac}$ is measured with the help of a current pre-amplifier. Three QPCs are defined with three pairs of constriction gates with voltage $V_\text{cg1}$, $V_\text{cg2}$ and $V_\text{cg3}$. The accumulation gate voltage $V_\text{ag}$ tunes the carrier density in the device globally. In QPC measurements, magnetic field is applied perpendicularly as well. \\

\section{Serial resistance subtraction}
As seen in Figure \ref{FigureS2}, QPCs are measured in a so-called two-terminal setup, where measurement circuit connect to the source (S) and drain (D) leads of the device. In this case, the measured resistance of the device is composed of three parts (1) contact resistance in the S/D leads, (2) resistance of the Ge quantum well between a dedicated QPC and two contacts, (3) resistance of a dedicated QPC. Normally, we believe contact resistance is constant even varying measurement conditions. However, the second part of the resistance is modulated with external magnetic field, especially in large magnetic fields. Therefore, the serial resistance $R_\text{s}$ (first two parts) need to be subtracted and the value should be adjusted depending on external magnetic field. 

In the presence of $R_\text{s}$, both differential conductance $G_\text{diff}$ and dc bias voltage $V_\text{b}$ need to be re-calculated. The differential conductance is calculated as $G_\text{diff}=1/(V_\text{ac}/I_\text{ac}-R_\text{s})$. In order to correct dc voltage bias, dc current through the device is required. Initially, we found that ac signal has a noticeable influence on dc current measurement. We therefore acquire the dc current by integrating ac signals at varied $V_\text{dc}$ with the formula $I_\text{dc}(V_\text{x})=\int_{0}^{V_\text{x}} \frac{I_\text{ac}}{V_\text{ac}} \,dV_\text{dc}$. After having $I_\text{dc}$, we correct voltage bias as $V_\text{b}=V_\text{dc}-I_\text{dc}\cdot R_\text{s}$.   

Serial resistance $R_\text{s}$ is chosen such that the first conductance plateau is $2e^2/h$ after subtraction. Considering the influence of magnetic field, $R_\text{s}$ is obtained and subtracted in such a way at each magnetic field.

\section{Additional data}
In this section, we provide additional figures where essential parameters displayed in the main article are extracted. 

\begin{figure*}[!t]
\centering
\includegraphics[width=0.98\linewidth]{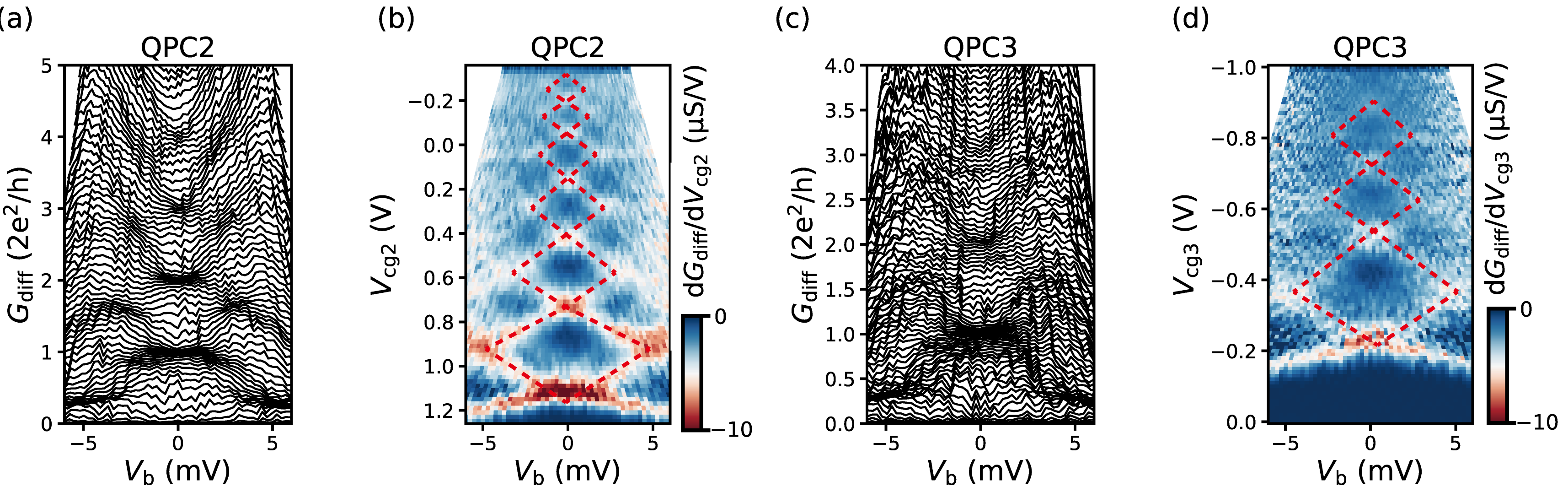}
\caption{(a), (b) Waterfall plot and transconductance graph of QPC2 at $V_\text{cg1}=V_\text{cg3}=-2.5\,\mathrm{V}$ and $B=0$. (c), (d) Waterfall plot and transconductance graph of QPC3 at $V_\text{cg1}=V_\text{cg2}=-2.5\,\mathrm{V}$ and $B=0$. Red dashed lines in transconductance graphs denote the boundaries of the diamonds, where quantization energies between successive 1D subbands are obtained. With Eq.(1) in the main text, the effective length $L_\text{n}$ of the first six subbands of QPC2 is estimated to be 12, 29, 42, 52, 65 and 80$\,\mathrm{nm}$, respectively. $L_\text{n}$ of the first three subbands of QPC3 is estimated to be 12, 29 and 42$\,\mathrm{nm}$, respectively.}\label{FigureS4}
\end{figure*}

Figure \ref{FigureS4} shows bias-spectroscopy measurements of QPC2 and QPC3 in the absence of magnetic field. From the figure, we obtain quantization energies of the 1D subbands for the two QPCs. Corresponding results are shown in Figure 2(d) in the main article. 

\begin{figure*}[!t]
\centering
\includegraphics[width=0.8\linewidth]{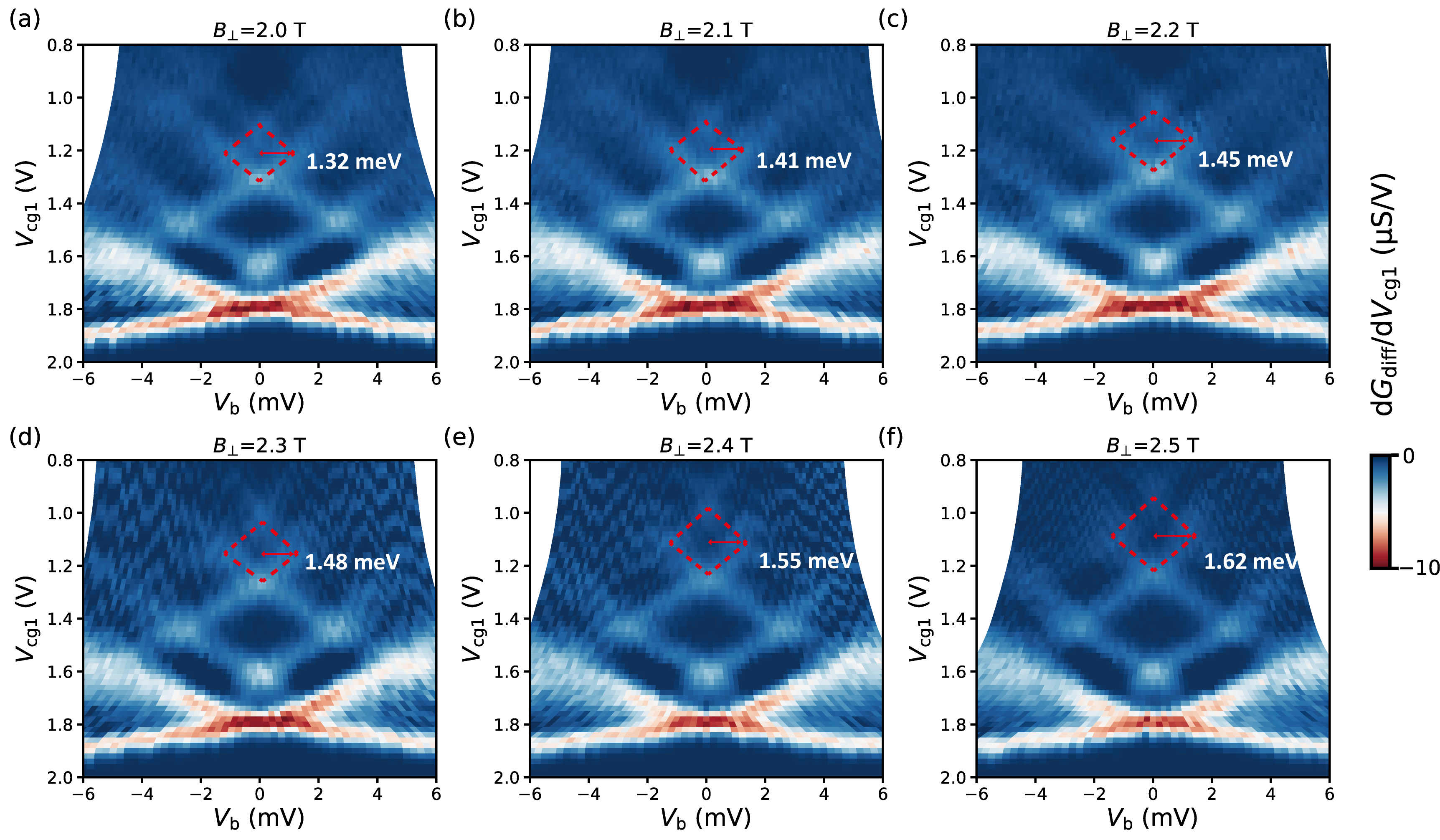}
\caption{Bias-spectroscopy measurements of QPC1 at different perpendicular magnetic fields $B_{\bot}$. Red dashed lines denote the diamond formed by the second subband with finite Zeeman energies. The extracted Zeeman energies of the second subband at different $B_{\bot}$ are displayed in Figure 3(e) of the main article.}\label{FigureS5}
\end{figure*}

\begin{figure*}[!t]
\centering
\includegraphics[width=0.8\linewidth]{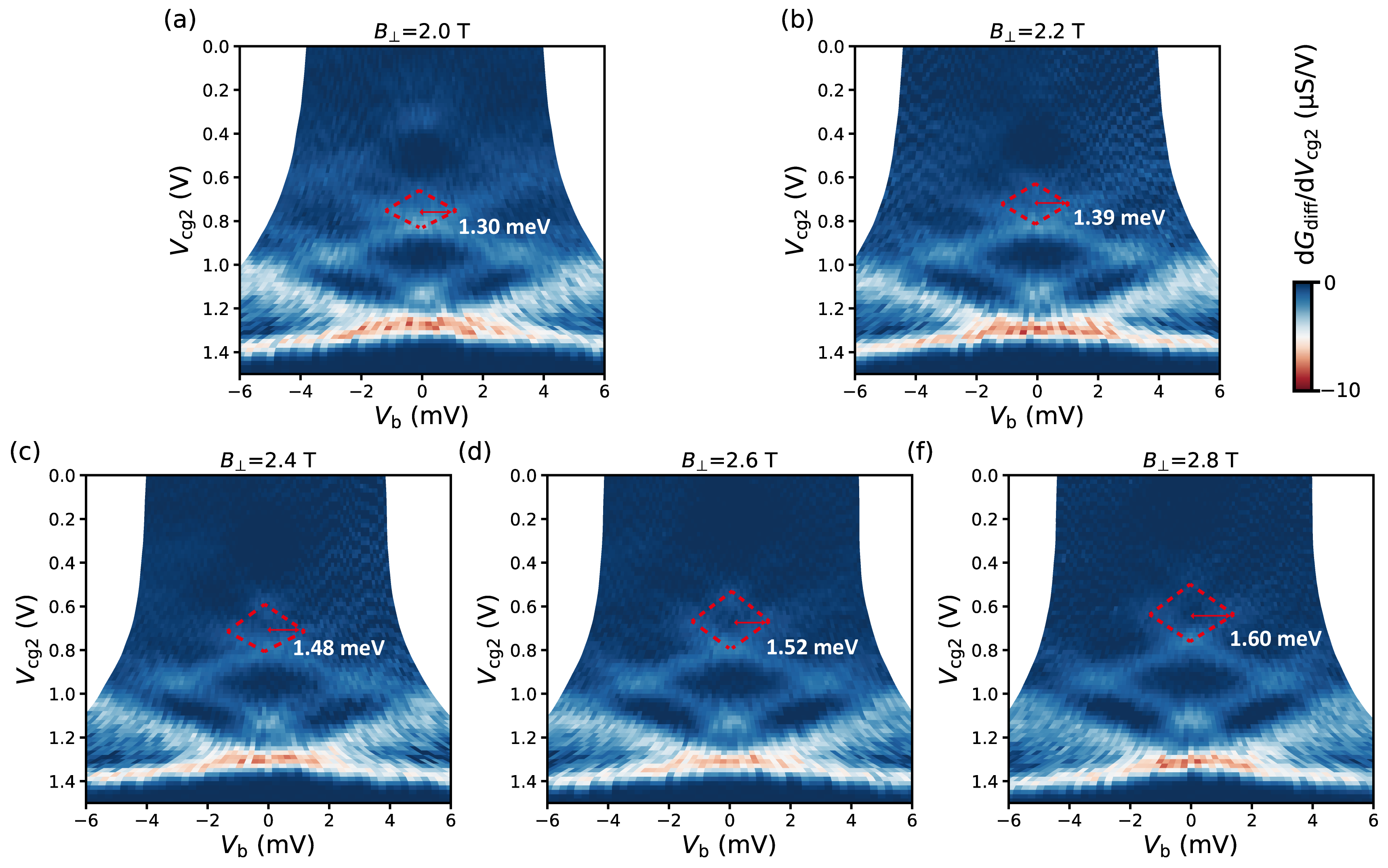}
\caption{Bias-spectroscopy measurements of QPC2 at different $B_{\bot}$. Red dashed lines denote the diamond formed by the second subband with finite Zeeman energies. The extracted Zeeman energies of the second subband at different $B_{\bot}$ are displayed in Figure 3(f) of the main article.}\label{FigureS6}
\end{figure*}

\begin{figure*}[!t]
\centering
\includegraphics[width=0.8\linewidth]{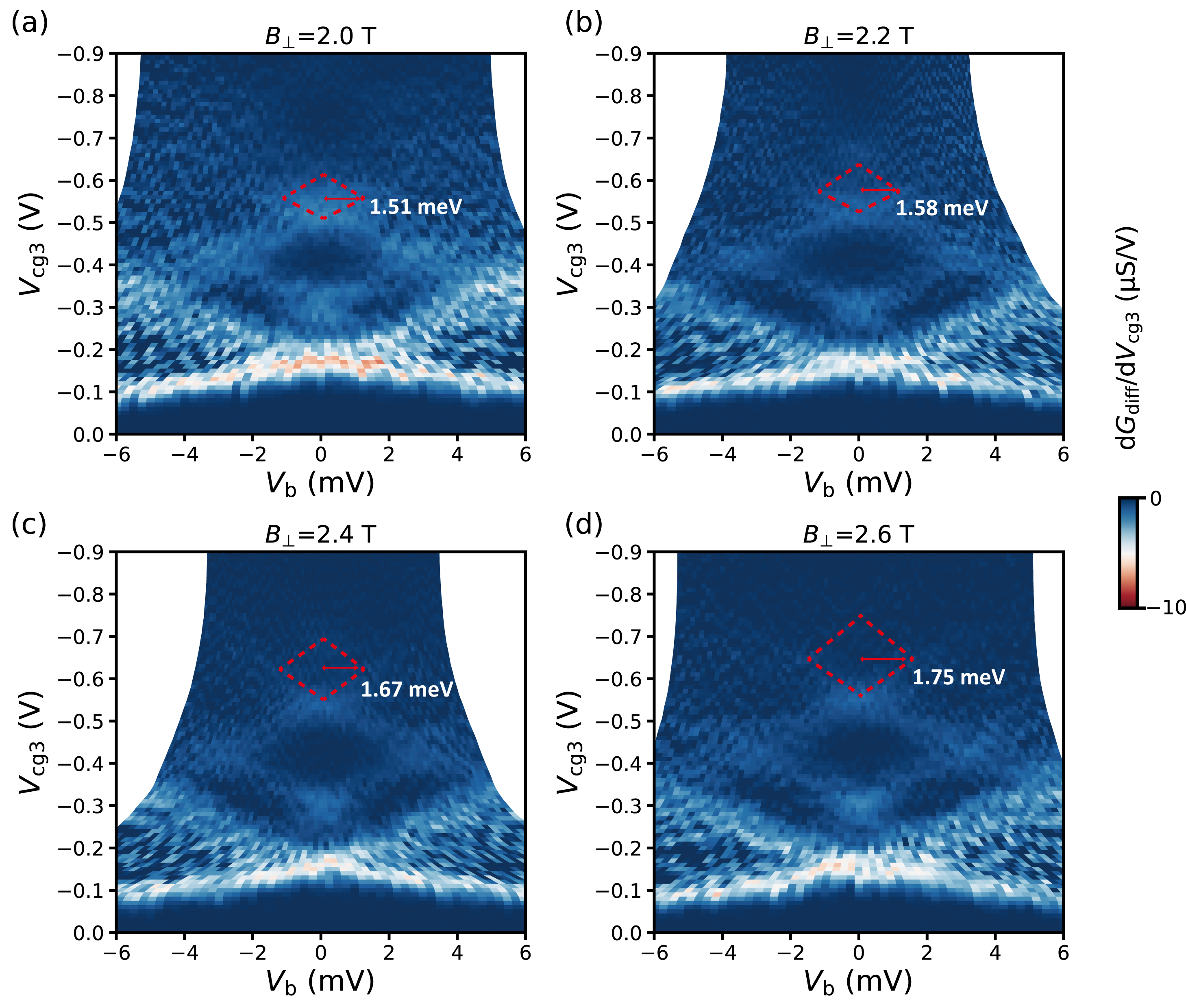}
\caption{Bias-spectroscopy measurements of QPC3 at different $B_{\bot}$. Red dashed lines denote the diamond formed by the second subband with finite Zeeman energies. The extracted Zeeman energies of the second subband at different $B_{\bot}$ are displayed in Figure 3(g) of the main article.}\label{FigureS7}
\end{figure*}

Figures \ref{FigureS5}-\ref{FigureS7} are bias-spectroscopy measurements of the three QPCs in perpendicular magnetic fields $B_{\bot}$. From these measurements, we obtain Zeeman energies of 1D subbands in QPCs at different $B_{\bot}$ and corresponding data are present in Figures 3(e)-3(g) in the main article.  

\bibliography{reference}% Produces the bibliography via BibTeX.